\documentclass[12pt]{iopart}

\usepackage[dvips]{graphicx,color}
\usepackage{psfrag}
\usepackage{pstricks}
\usepackage{subfigure}
\usepackage{iopams}

\newcommand{\fig}[1]{Figure$\,$\ref{#1}}
\newcommand{\eq}[1]{Eq.(\ref{#1})}

\begin{document}
\title[Characteristics of integrated magneto-optical traps for atom chips]{Characteristics of integrated magneto-optical traps for atom chips}
\author{S Pollock$^{1}$, J P Cotter$^{1}$, A Laliotis$^{1,2}$, F Ramirez-Martinez$^{1,3}$ and E A Hinds$^{1}$}
\address{$^{1}$ The Centre for Cold Matter, Blackett Laboratory, Imperial College London, SW7 2AZ}
\address{$^{2}$ Laboratoire de Physique des Lasers UMR 7538 du CNRS, Universit\'{e} Paris-13, F-93430, Villetaneuse, France}
\address{$^{3}$ LNE-SYRTE, Observatoire de Paris, UPMC, CNRS, 61 av de l'Observatoire, 75014 Paris, France}
\ead{j.cotter@imperial.ac.uk}

\begin{abstract}
We investigate the operation of pyramidal magneto-optical traps (MOTs) microfabricated in silicon. Measurements of the loading and loss rates give insight into the role of the nearby surface in the MOT dynamics. Studies of the fluorescence versus laser frequency and intensity allow us to develop a simple theory of operation.  The number of $^{85}$Rb atoms trapped in the pyramid is approximately $L^6$, where $L \lesssim 6$ is the size in mm. This follows quite naturally from the relation between capture velocity and size and differs from the $L^{3.6}$ often used to describe larger MOTs. Our results constitute substantial progress towards fully integrated atomic physics experiments and devices.
\end{abstract}

\maketitle

\section{Introduction}
Atom chips are made by micro-fabrication methods and are used to control, trap, and manipulate ultra-cold atoms \cite{Fortagh07,reichelbook}. Normally, cold atoms are prepared far from the surface by a reflection MOT \cite{reichel99,vangeleyn09} or other standard source then transferred to the microscopic traps of the atom chip through a delicate sequence of changes to laser beams and magnetic fields. These steps make it awkward to load a single trap and difficult to load several traps in parallel. We have recently demonstrated a simple alternative whereby small pyramids etched into a silicon wafer may be used together with integrated current-carrying wires to trap small atom clouds directly from a thermal vapour \cite{trupke06,lewis09,pollock09}. Compared with other atom-chip sources this configuration is very simple to operate, and because they are micro-fabricated it is easy to form several pyramid MOTs on the same chip.

In this paper we study the operation of integrated MOTs with $^{85}$Rb atoms. In Sec.~\ref{sec:setup}, we describe how the small trapped atom clouds are imaged, then in Sec.~\ref{sec:loadloss} we use these images to measure the loading and loss rates of the trap and hence to determine the capture velocity. Section~\ref{sec:positiondependence} discusses how the number of atoms is affected by the position of the cloud within the pyramid, leading to an understanding of how the walls influence the capture of atoms from the vapour and the loss of atoms that are already trapped. We note in Sec.~\ref{sec:intensitydependence} that the position of the cloud depends on the laser intensity and we measure that effect. Drawing on these results, we are able in Sec.~\ref{sec:understanding} to understand how the MOT fluorescence signal optimises with respect to laser detuning and intensity. Section~\ref{subsec:ScalingLaws} demonstrates that the number of atoms in the cloud scales in a simple way as the size of the pyramid is changed. Finally, in Sec.~\ref{sec:summary} we summarise our results and briefly discuss future prospects for integrated MOTs.

\section{\label{sec:setup}Experimental Setup}
\begin{figure}[t]
\centering
\subfigure[The wafer\label{fig:wafer}]{\includegraphics[width = 0.45\columnwidth]{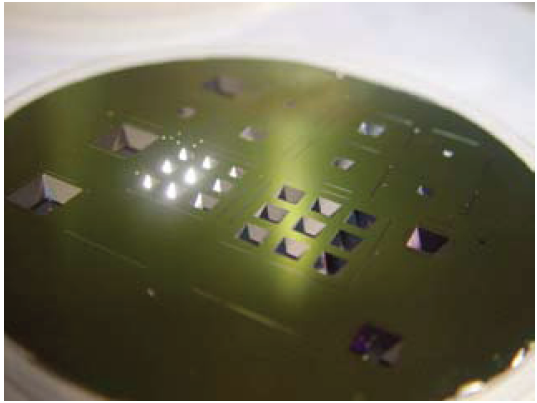}}
\hspace{1cm}
\subfigure[The chip package\label{fig:ChipPackage}]{\includegraphics[width = 0.45\columnwidth]{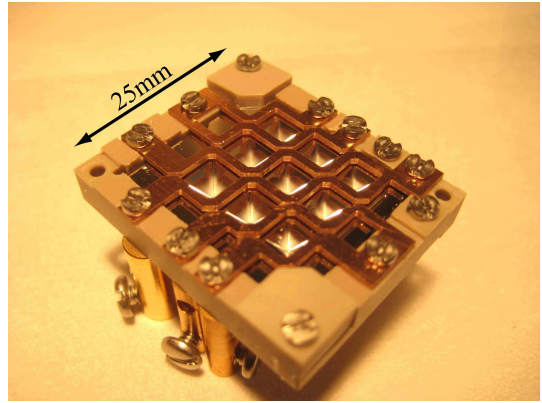}}
\caption{(a) Pyramids etched into a silicon wafer. The wafer is $3$mm thick and $4^{\prime\prime}$ in diameter.
(b) A $3 \times 3$ array of silicon pyramids mounted in PEEK holder $25 \times 30\,$mm$^{2}$. The magnetic field required for trapping atoms is provided by a zig-zag array of copper wires above and below the pyramids.
}
\end{figure}

\fig{fig:wafer} shows a $3$\,mm-thick, $4^{\prime\prime}$-diameter silicon wafer, cut along the $\left[100\right]$ crystal plane, which we have etched in potassium hydroxide to form concave pyramidal hollows. The sides of the pyramids range in length $L$ from $4.2\,$mm down to $1.3\,$mm, while the apex angle $\theta = 2\arctan (1/\sqrt{2})=70.5^\circ$ is fixed by the crystal structure. Each set of pyramids released from the wafer is coated with $25\,$nm of aluminium to achieve a reflectivity of $75\%$, so that the pyramid MOTs can work without the need to mask any part of the surface \cite{lewis09}. Initial tests were performed on an isolated pyramid, with hand-wound coils generating the trapping magnetic field. This was later replaced by the chip package shown in \fig{fig:ChipPackage} containing a $3 \times 3$ array of pyramids. A carrier machined from PEEK supports both the pyramid array and the current-carrying copper structure that generates the required array of magnetic quadrupoles. Typical quadrupole field gradients used for this package are $0.3\,$Tm$^{-1}$, produced by a current of $3\,$A. The assembly is mounted in an ultra high vacuum chamber with the pyramid openings facing downwards. Rubidium atoms are provided by a current-activated dispenser. More details relating to the fabrication of these pyramids as well as the chip package and laser system are given in previous publications \cite{trupke06,lewis09,pollock09}.

\begin{figure}[b]
\centering
\subfigure[Before image filtering\label{fig:img1}]{\includegraphics[width = 0.45\columnwidth]{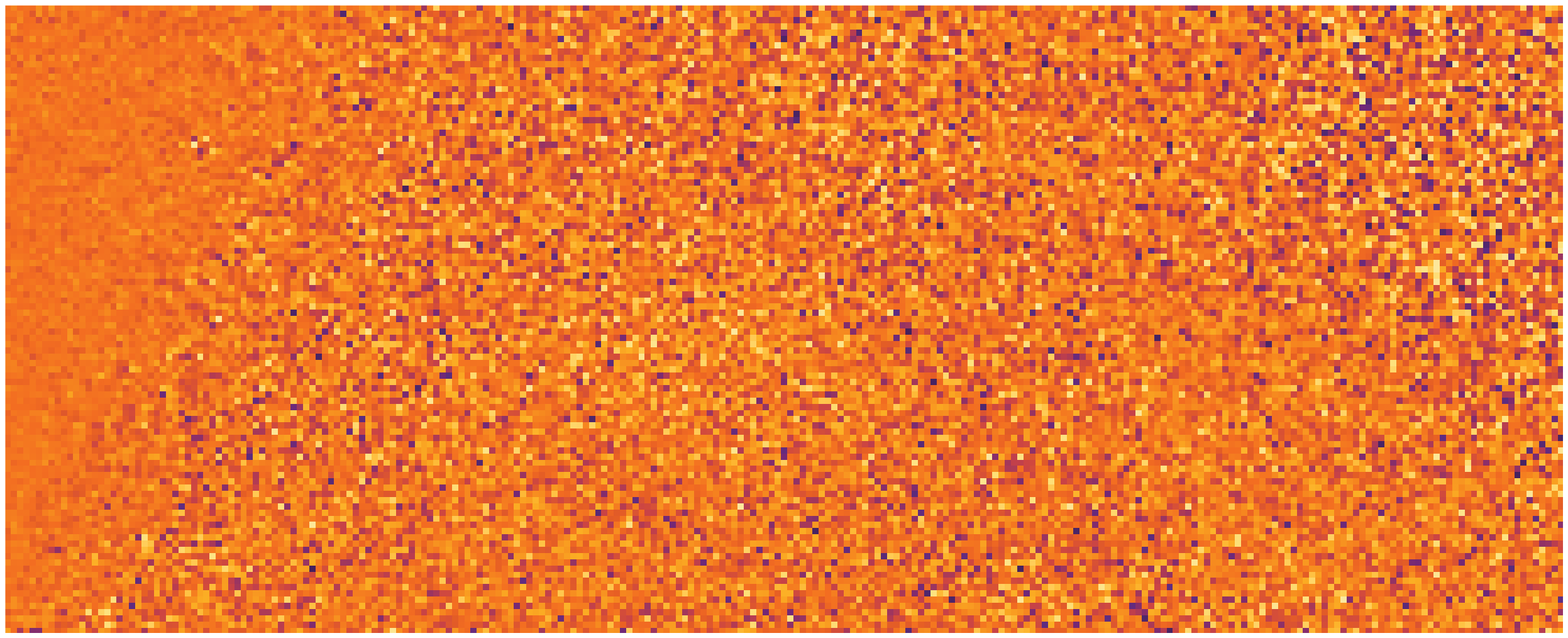}}
\hspace{1cm}
\subfigure[After image filtering\label{fig:img2}]{\includegraphics[width = 0.45\columnwidth]{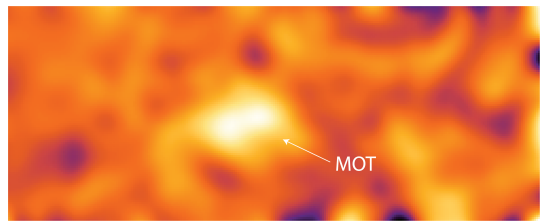}}
\caption{
A cloud of $7\times 10^{3}$ atoms with Gaussian radius $\sigma_{MOT}\simeq100\,\mu$m. a) Average of 500 background-subtracted images. The MOT is not clearly visible here against the background noise. b) Gaussing filtering of this image clearly reveals the MOT.
\label{fig:imgB}}
\end{figure}

Atoms trapped in the pyramids are detected by imaging their fluorescence onto a CCD camera. We use a combination of background subtraction and image filtering to distinguish MOT fluorescence from the light scattered by the pyramid surfaces and by untrapped rubidium vapour. The MOT is turned on and off by a uniform magnetic field of $100\,$mG along the pyramid axis. This moves the magnetic quadrupole to a position where a MOT is not supported, while making almost no change to the fluorescence from untrapped atoms. Since the noise in the background-subtracted images is dominated by photon shot noise, we are able to improve the visibility of the MOT by averaging many images. Figure~2 shows a cloud of $7 \times 10^{3}\,$atoms. Although we have averaged over 500 exposures of $40\,$ms each, the MOT is difficult to distinguish from the background in \fig{fig:img1}. The visibility is much improved in \fig{fig:img2} by convolving \fig{fig:img1} with a Gaussian \cite{book_MedImg} having a standard deviation of $5\,$pixels ($45\,\mu$m), which is a little smaller than the $\sim100\,\mu$m radius of the MOT. In this way, we can discern MOTs as small as $200\,$atoms at the moment and we anticipate even better sensitivity in future pyramids with smoother faces \cite{pollock09}.

\section{Results}
\subsection{Loading and loss rates \label{sec:loadloss}}
\begin{figure}[b]
{\footnotesize
\subfigure[Loading curves\label{fig:loadingcurves}]{
\begin{psfrags}
\psfrag{Time}[tc][tc]{t (s)}%
\psfrag{N}[bc][tc]{Number of atoms}%
\psfrag{0.0}[tc][tc]{$0.0$}%
\psfrag{0.2}[tc][tc]{$0.2$}%
\psfrag{0.4}[tc][tc]{ $0.4$}%
\psfrag{0.6}[tc][tc]{ $0.6$}%
\psfrag{0.8}[tc][tc]{ $0.8$}%
\psfrag{1.0}[tc][tc]{ $1.0$}%
\psfrag{0}[cr][cr]{ $0$}%
\psfrag{500}[cr][cr]{ $500$}%
\psfrag{1000}[cr][cr]{ $1000$}%
\psfrag{1500}[cr][cr]{ $1500$}%
\psfrag{2000}[cr][cr]{ $2000$}%
\psfrag{2500}[cr][cr]{ $2500$}%
\psfrag{3000}[cr][cr]{ $3000$}%
\psfrag{Rb}[cr][cr]{ $n_{vap} ({\tiny m^{-3}})$}%
\psfrag{A}[cr][cr]{ \tiny $1.3\times 10^{13}$}%
\psfrag{B}[cr][cr]{ \tiny $2.7\times 10^{13}$}%
\psfrag{C}[cr][cr]{ \tiny $3.8\times 10^{13}$}%
\psfrag{D}[cr][cr]{ \tiny $5.4\times 10^{13}$}%
\psfrag{E}[cr][cr]{ \tiny $7.7\times 10^{13}$}%
\psfrag{F}[cr][cr]{ \tiny $10.9\times 10^{13}$}%
\psfrag{G}[cr][cr]{ \tiny $19.5\times 10^{13}$}%
\psfrag{H}[cr][cr]{ \tiny $27.6\times 10^{13}$}%
\psfrag{I}[cr][cr]{ \tiny $41.0\times 10^{13}$}%
\includegraphics[height=5.5cm]{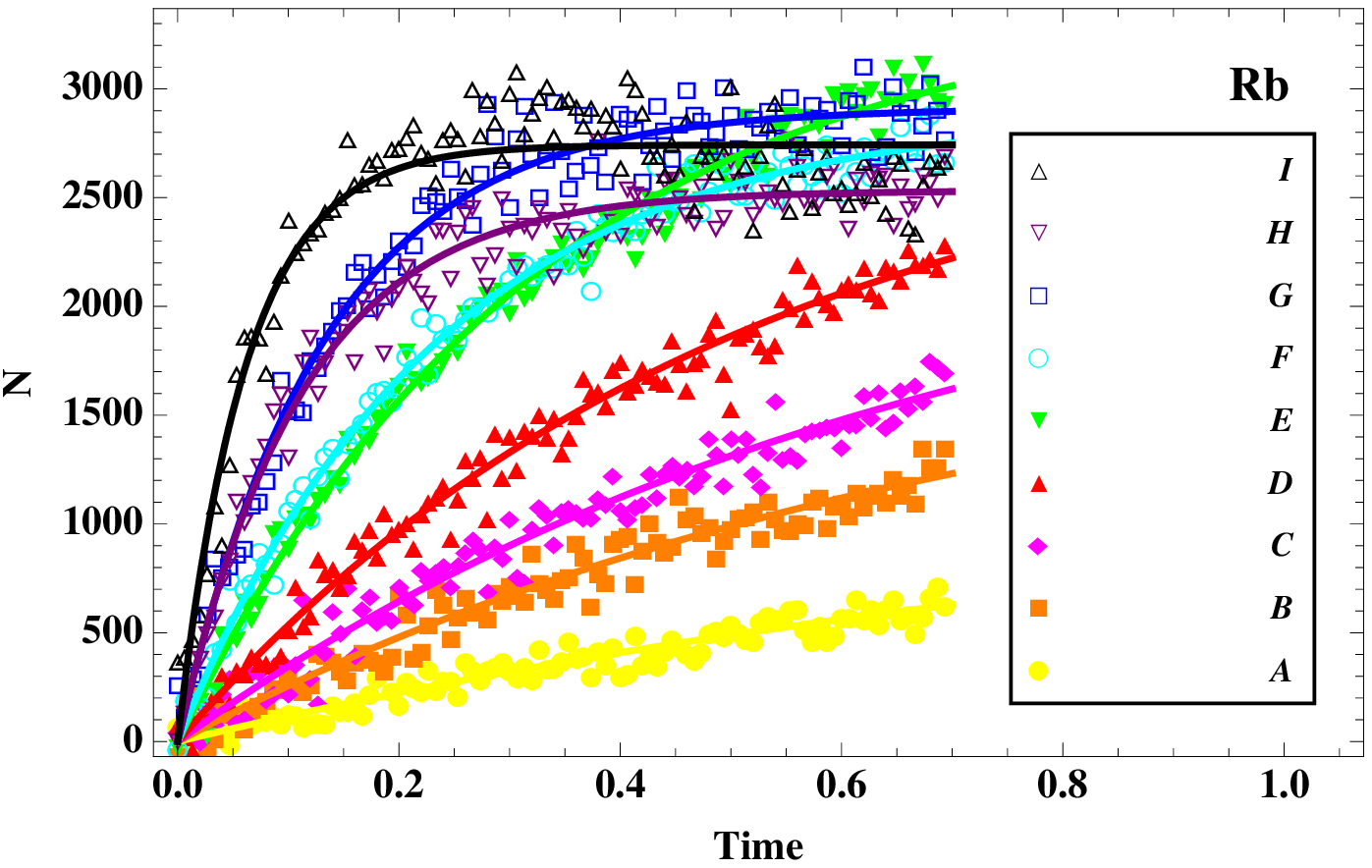}
\end{psfrags}
}
\hspace{1cm}
\subfigure[Loss and capture rates.\label{fig:loadingloss1}]{
\begin{psfrags}
\psfrag{A}[tc][tc]{$15$}%
\psfrag{B}[tc][tc]{$5$}%
\psfrag{C}[tc][tc]{$40$}%
\psfrag{D}[tc][tc]{$20$}%
\psfrag{E}[tc][tc]{$1$}%
\psfrag{F}[tc][tc]{$2$}%
\psfrag{G}[tc][tc]{$3$}%
\psfrag{H}[tc][tc]{$4$}%
\psfrag{X}[tc][tc]{n$_{vap}$ ($10^{14}$m$^{-3}$)}%
\psfrag{Q}[tc][tc]{$1/\tau$~(s$^{-1}$)}%
\psfrag{Z}[tc][tc]{$R~(10^{3}\,$s$^{-1})$}%
\includegraphics[height=5.6cm]{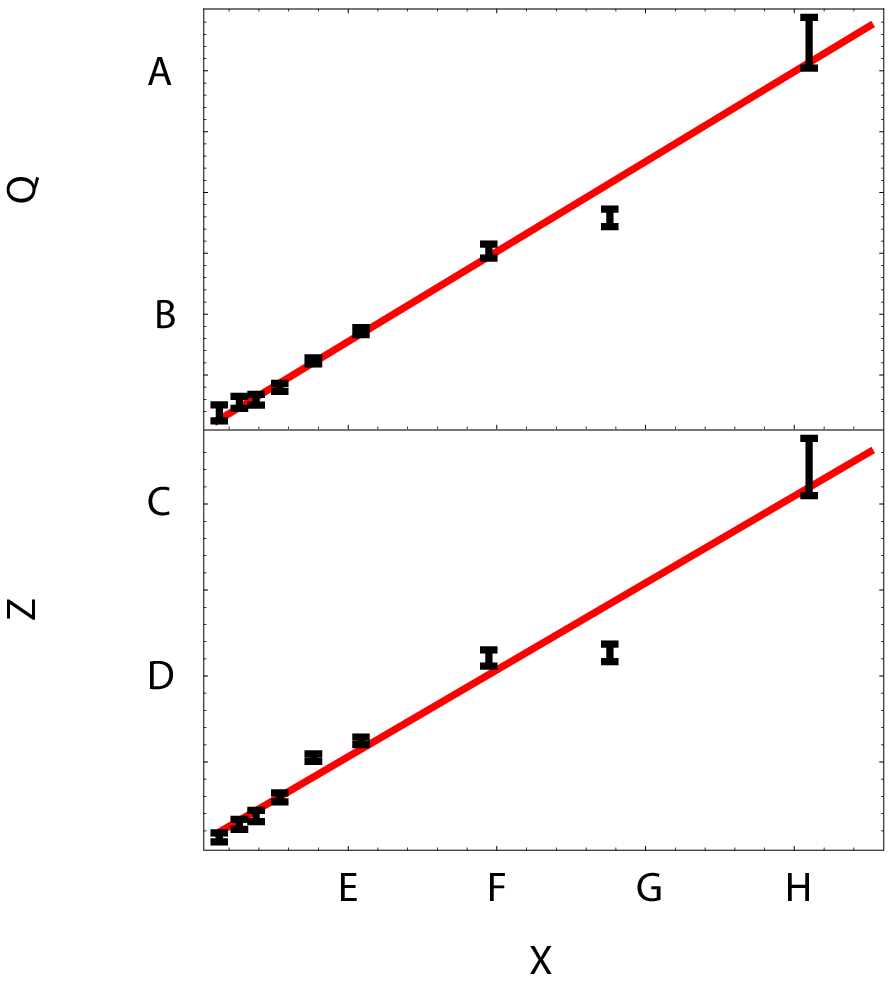}
\end{psfrags}
}
\caption{a) Number of atoms in the MOT $N(t)$ versus time for several values of the background rubidium vapour density $n_{vap}$. The lines are least square fits of \eq{eq:load} to each data set. b) Loss rate $1/\tau$ and capture rate $R$ both exhibit a linear dependance on rubidium pressure. Linear least-squares fits (solid lines) give $1/\tau = (3.5(2) \times10^{-14} n_{vap} + 0.3(2))\,$s$^{-1}$ and $R = (9.6(8) \times 10^{-11} n_{vap} + 1(1)10^{3})\,$atoms.s$^{-1}$. The numbers in parentheses are standard errors given by the fits.
\label{fig:loadingcurves2}}
}
\end{figure}
The number of atoms in the trap $N(t)$ depends on the balance between the rate $R$ of capture from the background Rb vapour and the rate $1/\tau$ at which atoms are lost from the MOT by collisions with the background gas. The loss due to collisions between trapped atoms can be neglected for the traps considered here. Thus the number of trapped atoms is well approximated by
\begin{equation}
N(t)=N_{\infty}(1-e^{-t/\tau}),
\label{eq:load}
\end{equation}
where $N_{\infty} = R\tau$ is the steady-state number of atoms. The background Rb pressure is adjusted by changing the current in the dispenser. We determine the corresponding density and temperature of the Rb vapour, $n_{vap}$ and $T_{vap}$, by passing a probe laser through the chamber and measuring the optical thickness at resonance and the Doppler width of the absorption. \fig{fig:loadingcurves} shows $N(t)$ measured at several values of $n_{vap}$ in a pyramid of side length $L=4.2\,$mm. At each density, we turned the trap on at $t=0$ and recorded $500$ MOT images, equally spaced over an interval of $3.3\,$s. On fitting Eq.~(\ref{eq:load}) to these measurements, we obtain the loss and capture rates plotted in \fig{fig:loadingloss1}. One sees that the loss rate is proportional to the Rb vapour density, as  expected from the relation $1/\tau = n_{vap} \sigma_{loss}\bar{v}_{vap}$, where $\sigma_{loss}$ is the cross section for a trapped atom to be kicked out of the trap by collision with a hot Rb atom and $\bar{v}_{vap}$ is the mean speed of atoms in the vapour. At zero background pressure of rubidium, $n_{Rb} = 0$, our fit to the data in \fig{fig:loadingloss1} gives a loss rate of $0.3(2)\,$s$^{-1}$, which is consistent with zero. This shows that although the walls of the pyramid are close to the  trapped atoms, any loss to the walls is nevertheless very slow in comparison with the loss rate from collisions with the background rubidium vapour. For the measured temperature $T_{vap}=298\,$K, the mean speed is $272\,$m/s, which yields a measured value for the trap-loss cross section of $\sigma_{loss} = 1.29(7)\times 10^{-16}\,{\rm m}^2$. This cross section is related to the velocity needed to escape from the trap, given by~\cite{steane92} $\sigma_{loss} = \pi \left( 4 C_{3}/m v_{esc} \bar{v}_{vap} \right)^{2/3}$, where   $C_{3} = 5.8 \times 10^{-48}\,$ Jm$^{3}$ is the coefficient of the dominant dipole-dipole interaction and $m$ is the mass of the $^{85}$Rb atom. Hence the escape velocity is $v_{esc} = 2.3(2)\,{\rm m s}^{-1}$.

The measured capture rate $R$, shown in \fig{fig:loadingloss1}, is also proportional to $n_{vap}$, with a slope of $9.6(8)\times 10^{-11}{\rm m}^3 {\rm s}^{-1}$. In a simple model for this rate, assuming a spherical stopping region of cross sectional area $A$  (see Appendix), this slope is to equal $8 A v_c^4/(3\pi^2\bar{v}_{vap}^3)$, where $v_c$ is the maximum velocity of atoms that are captured. We can reasonably assume that this capture velocity is twice the escape velocity since the distance available for capture is approximately twice the distance needed for escape and the friction force is proportional to velocity. Taking $v_c=2v_{esc}=4.6\,{\rm m s}^{-1}$, we obtain a capture area of $A=16\,\rm mm^2$, which is reassuringly close to the physical opening area of our $L=4.2\,\rm mm$ pyramid. Henceforth we will take it that $A\simeq L^2$. Combining our results for the capture and loss rates we obtain an expression for the steady state number of atoms in the MOT,
\begin{equation}
N_{\infty} =R \tau\simeq \frac{8 L^2}{3 \pi^2 \sigma_{loss}}\frac{v_{c}^{4}}{\bar{v}_{vap}^{4}} \,.
\label{eq:N}
\end{equation}
Since we expect the capture velocity to be proportional to the size of the MOT, i.e. $v_c\propto L$, the number of trapped atoms should scale with pyramid size in proportion to $L^6$, a relation that we verify in Sec.~\ref{subsec:ScalingLaws}.

\subsection{\label{sec:positiondependence}Position dependent atom number in the MOT}

\begin{figure}[t]
{\footnotesize
\centering
\subfigure[Variation of MOT fluorescence]{
\begin{psfrags}
\psfrag{BoxBxG}[tc][tc]{\scriptsize$B_{x}$ (G)}%
\psfrag{BoxByG}[bc][bc]{\scriptsize$B_{y}$ (G)}%
\psfrag{dmmA}[Bc][Bc]{\scriptsize$d_{x}$ (mm)}%
\psfrag{dmm}[bc][tc]{\scriptsize$d_{y}$ (mm)}%
\psfrag{EStyleFormA}[cl][cl]{~~~~$0.0$}%
\psfrag{EStyleFormB}[cl][cl]{~~~~$1.3$}%
\psfrag{EStyleForm}[cl][cl]{ $-1.3$}%
\psfrag{NStyleFormA}[bc][bc]{ $0.0$}%
\psfrag{NStyleFormB}[bc][bc]{ $1.3$}%
\psfrag{NStyleForm}[bc][bc]{ $-1.3$}%
\psfrag{S0}[tc][tc]{$0.0$}%
\psfrag{S11}[tc][tc]{ $1.0$}%
\psfrag{S5}[tc][tc]{~}%
\psfrag{Sm11}[tc][tc]{ $-1.0$}%
\psfrag{Sm5}[tc][tc]{~}%
\psfrag{W0}[cr][cr]{$0.0$}%
\psfrag{W11}[cr][cr]{ $1.0$}%
\psfrag{W5}[cr][cr]{~}%
\psfrag{Wm11}[cr][cr]{ $-1.0$}%
\psfrag{Wm5}[cr][cr]{ ~}%
\includegraphics[height=5cm]{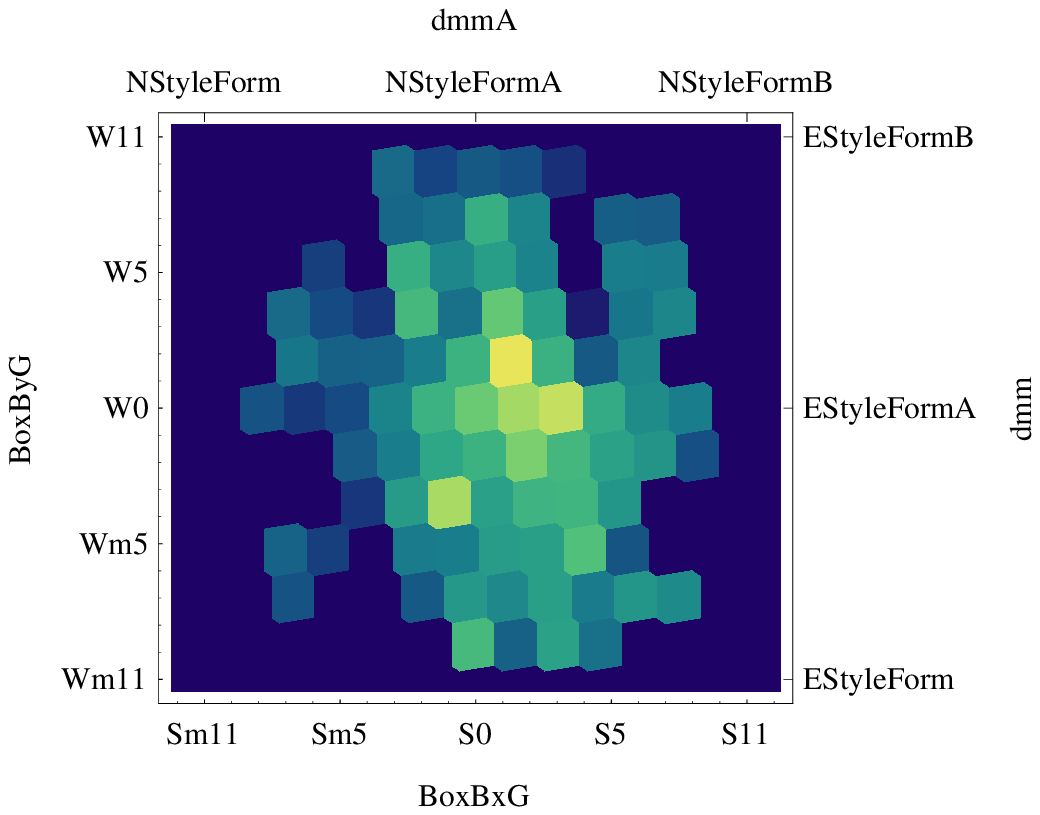}
\end{psfrags}
}
\hspace{1cm}
\subfigure[Loss rate to the surface]{
\begin{psfrags}
\psfrag{X1}[cc][cc]{$d$ (mm)}%
\psfrag{X2}[cc][cc]{\scriptsize $d^{2}$ (mm$^{2}$)}%
\psfrag{Y1}[cB][cl]{$1/\tau$ (s$^{-1}$)}%
\psfrag{Y2}[cc][cc]{\scriptsize $1/\tau - 1/\tau_{0}$}%
\psfrag{5}[cr][cr]{~}%
\psfrag{10}[cr][cr]{10}%
\psfrag{15}[cr][cr]{~}%
\psfrag{20}[cr][cr]{20}%
\psfrag{25}[cr][cr]{~}%
\psfrag{30}[cr][cr]{30}%
\psfrag{35}[cr][cr]{~}%
\psfrag{40}[cr][cr]{40}%
\psfrag{0.2}[cr][cr]{0.2}%
\psfrag{0.3}[cr][cr]{0.3}%
\psfrag{0.4}[cr][cr]{0.4}%
\psfrag{0.5}[cr][cr]{0.5}%
\psfrag{0.6}[cr][cr]{0.6}%
\psfrag{0.7}[cr][cr]{0.7}%
\psfrag{0.5}[cr][cr]{0.5}%
\psfrag{0.11}[cr][cr]{\scriptsize 1}%
\psfrag{2.3}[cr][cr]{\scriptsize 10}%
\psfrag{3.9}[cr][cr]{\scriptsize 50}%
\psfrag{0.02}[cr][cr]{\scriptsize 0.02}%
\psfrag{0.04}[cr][cr]{~}%
\psfrag{0.06}[cr][cr]{\scriptsize 0.06}%
\psfrag{0.08}[cr][cr]{~}%
\psfrag{0.1}[cr][cr]{\scriptsize 0.10}%
\psfrag{0.12}[cr][cr]{~}%
\psfrag{0.14}[cr][cr]{\scriptsize 0.14}%
\psfrag{0.16}[cr][cr]{~}%
\includegraphics[height=5cm]{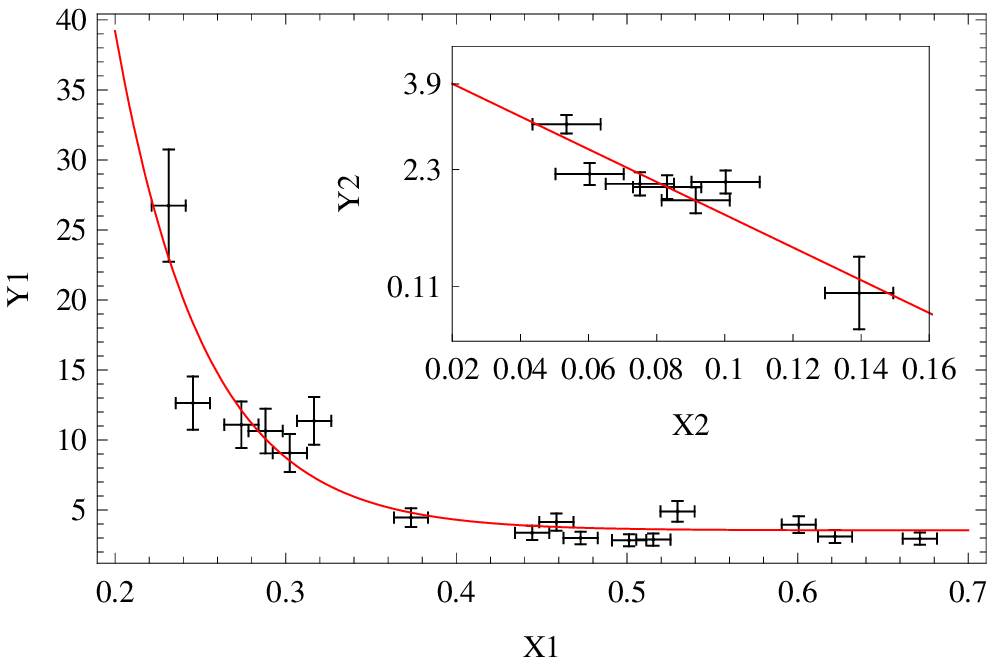}
\end{psfrags}
}
\caption{a) Variation of MOT fluorescence as the trap is translated horizontally within the pyramid by changing a bias magnetic field. The fluorescence decrease when the cloud approaches the walls b) Change of loss rate $1/\tau$ as a MOT of rms radius $120\,\mu$m approaches a surface. Data: measured rates derived from loading curves. Line: Fit to the function $A e^{-d^2/(2\sigma^2)}+1/\tau_0$. Fit parameter $\sigma=119\,\mu$m agrees with the size of the cloud. Over the same range of distances there is no discernable variation in the capture rate R. Inset: Semi-log plot of the same data.
\label{fig:HorizMove}}
}
\end{figure}

We are able to adjust the position of the MOT within the pyramid by applying a magnetic shim field. \fig{fig:HorizMove}(a) shows a map of the MOT fluorescence versus the shim field over a $2.6\,$mm square region in the horizontal plane $1.8\,$mm from the apex.  When the MOT approaches the wall of the pyramid we find that the fluorescence decreases quite abruptly due to a decrease of $N_{\infty}$.

In order to determine whether this is due to a decrease in $\tau$ or reduction of $R$ we set up an auxiliary experiment using a larger, aluminium-coated glass pyramid of the same apex angle in which we could reliably determine the distance between the MOT and a pyramid face by measuring the distance between the MOT and its reflection. The rms radius of this cloud was $120(10)\,\mu$m. Additional loading-curve measurements allowed us to determine the loading and loss rates separately, just as in Section \ref{sec:loadloss}. Over the region where the MOT disappears, the loading rate is constant, whereas the loss rate exhibits a dramatic increase. The loss rate is plotted in \fig{fig:HorizMove}(b) {\em versus} the distance $d$ from the centre of the MOT to the surface. The data show loss to the surface in addition to a background collisional rate $1/\tau_{0}$.  A good fit is obtained with the function $1/\tau=Ae^{-d^2/(2\sigma^2)}+1/\tau_0$. This form is motivated by noting that the flux of MOT atoms crossing a plane at distance $d$ from the centre is proportional to $e^{-d^2/(2\sigma^2)}$, when the MOT density is a Gaussian of rms radius $\sigma$. The fit in \fig{fig:HorizMove}(b) yields the value $\sigma=119\,\mu$m, in agreement with the measured cloud size. We conclude that atoms are lost when the MOT approaches a surface because those at the edge of the cloud collide with it. This attenuates the MOT when the cloud comes within within a few hundred  $\mu$m of the surface.

\begin{figure}[b]
\centering
{\footnotesize
\subfigure{
\begin{psfrags} 
\psfrag{A}[cc][cc]{\textcolor{white}{\tiny$1.50$}}%
\psfrag{B}[cc][cc]{\textcolor{white}{\tiny$1.56$}}%
\psfrag{C}[cc][cc]{\textcolor{white}{\tiny$1.64$}}%
\psfrag{D}[cc][cc]{\textcolor{white}{\tiny$1.71$}}%
\psfrag{E}[cc][cc]{\textcolor{white}{\tiny$1.78$}}%
\psfrag{F}[cc][cc]{\textcolor{white}{\tiny$1.86$}}%
\psfrag{G}[cc][cc]{\textcolor{white}{\tiny$1.93$}}%
\psfrag{H}[cc][cc]{\textcolor{white}{\tiny$2.00$}}%
\psfrag{I}[cc][cc]{\textcolor{white}{\tiny$2.07$}}%
\psfrag{J}[cc][cc]{\textcolor{white}{\tiny$2.14$}}%
\psfrag{K}[cc][cc]{\textcolor{white}{\footnotesize Distance from the apex (mm)}}%
\psfrag{L}[cc][cc]{\textcolor{white}{\tiny$1$\,mm}}%
\psfrag{M}[cc][cc]{\textcolor{white}{\large(a)}}%
\includegraphics[height=3.7cm]{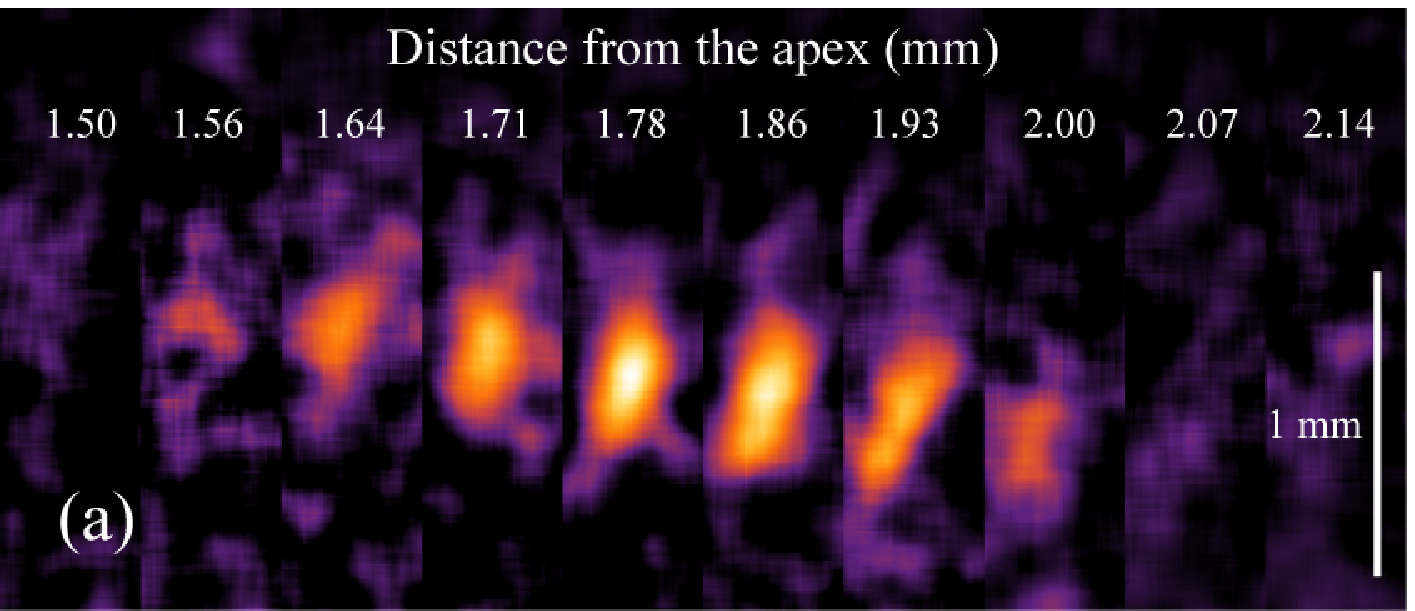}
\end{psfrags}
}
\hspace{0.25cm}
\subfigure{
\begin{psfrags}
\psfrag{a}[tc][tc]{ \large (b)}%
\psfrag{S11}[tc][tc]{ $1.0$}%
\psfrag{S121}[tc][tc]{ $1.2$}%
\psfrag{S141}[tc][tc]{ $1.4$}%
\psfrag{S161}[tc][tc]{ $1.6$}%
\psfrag{S181}[tc][tc]{ $1.8$}%
\psfrag{S21}[tc][tc]{ $2.2$}%
\psfrag{S221}[tc][tc]{ $2.2$}%
\psfrag{S241}[tc][tc]{ $2.4$}%
\psfrag{W0}[cr][cr]{ $0$}%
\psfrag{W11}[cr][cr]{ $1.0$}%
\psfrag{W2}[cr][cr]{ $0.2$}%
\psfrag{W4}[cr][cr]{ $0.4$}%
\psfrag{W6}[cr][cr]{ $0.6$}%
\psfrag{W8}[cr][cr]{ $0.8$}%
\psfrag{x}[tc][tc]{ Distance from apex (mm)}%
\psfrag{y}[bc][bc]{ Signal (Arb. Un.)}%
\includegraphics[height=4cm]{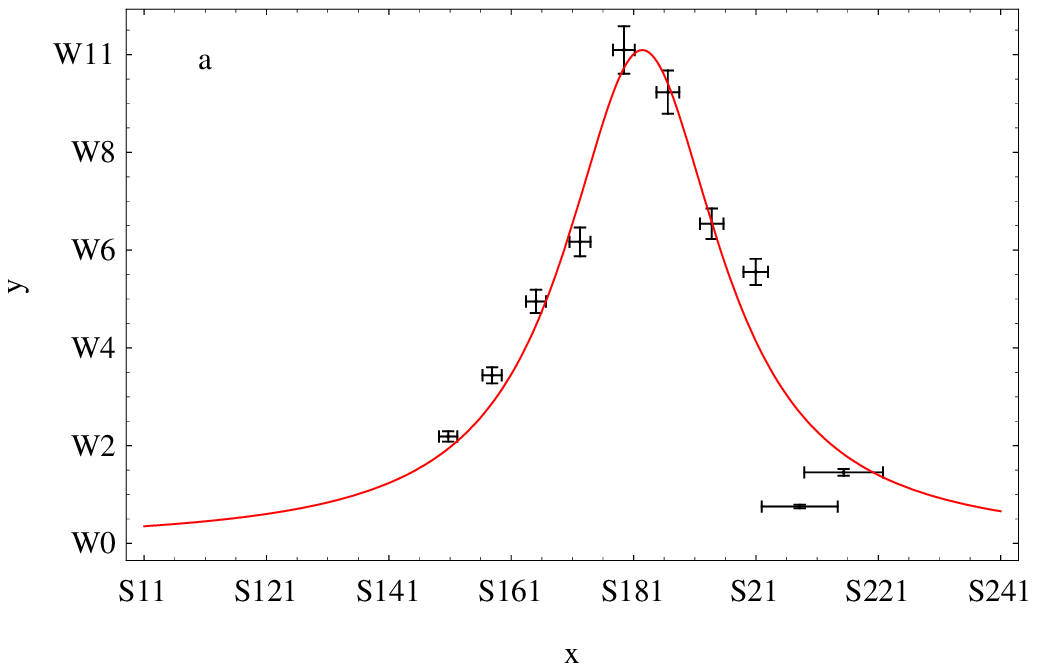}
\end{psfrags}
}
}
\caption{(a) Images of the MOT at various distances from the pyramid apex. The cloud height is controlled by moving the centre of the magnetic quadrupole using a vertical magnetic field. The images are compressed horizontally for the purpose of this display.  b) Data points: MOT fluorescence integrated over the image and normalised to the largest signal. Line: Lorentzian to guide the eye.
\label{fig:VertMove}}
\end{figure}

The bias magnetic field can also move the MOT up and down along the axis of the pyramid. \fig{fig:VertMove}(a) shows a series of MOT images taken at different distances from the apex of a $L=4.2\,$mm pyramid. The integrated MOT signals are plotted versus distance from the apex in \fig{fig:VertMove}(b). We see that the MOT is brightest at distances in the range $1.6-2.0\,$mm, centred roughly on $0.43L$. Further away from the apex, the capture rate $R$ decreases because the MOT beams are cut off by the aperture of the pyramid, the last ray (called type-1 in \cite{trupke06}) crossing the axis at $3L/\sqrt{32}=2.2\,$mm from the apex. Beyond this distance a MOT is unable to form.  The number of atoms also decreases as the MOT moves closer to the apex. This decrease turns on too soon to be explained by direct loss of atoms to the surfaces of the pyramid. Instead, we think it may be due to a drop in the capture velocity, which causes a decrease in the capture rate $R$. In numerical simulations, we see that atoms being captured need to explore a volume as they are slowed down and the available volume is restricted near the apex of the pyramid (compared with a plane surface at the same distance), thereby reducing the capture velocity. The over-all behaviour of the points in \fig{fig:VertMove}(b) is conveniently summarised by a Lorentzian function fitted to the data.

\subsection{\label{sec:intensitydependence}Intensity-dependent equilibrium position of the MOT}

\begin{figure}[b]
\centering
{\footnotesize
\begin{psfrags}
\psfrag{S0}[tc][tc]{ $~$}%
\psfrag{S122}[tc][tc]{ $12$}%
\psfrag{S12}[tc][tc]{ $~$}%
\psfrag{S142}[tc][tc]{ $~$}%
\psfrag{S21}[tc][tc]{ $~$}%
\psfrag{S41}[tc][tc]{ $4$}%
\psfrag{S61}[tc][tc]{ $~$}%
\psfrag{S81}[tc][tc]{ $8$}%
\psfrag{W0}[cr][cr]{ $0.0$}%
\psfrag{W11}[cr][cr]{ $1.0$}%
\psfrag{W151}[cr][cr]{ $~$}%
\psfrag{W161}[cr][cr]{ $1.6$}%
\psfrag{W1651}[cr][cr]{ $~$}%
\psfrag{W171}[cr][cr]{ $1.7$}%
\psfrag{W1751}[cr][cr]{ $~$}%
\psfrag{W181}[cr][cr]{ $1.8$}%
\psfrag{W1851}[cr][cr]{ $~$}%
\psfrag{W191}[cr][cr]{ $1.9$}%
\psfrag{W1951}[cr][cr]{ $~$}%
\psfrag{W21}[cr][cr]{ $2.0$}%
\psfrag{W5}[cr][cr]{ $~$}%
\psfrag{X2}[tc][tc]{ ~}
\psfrag{X}[tc][tc]{ $I$ (mW\,cm$^{-2}$)}%
\psfrag{Y2}[bc][bc]{ ~}%
\psfrag{Y}[bc][bc]{ distance to apex (mm)}%
\includegraphics[width=0.5\columnwidth]{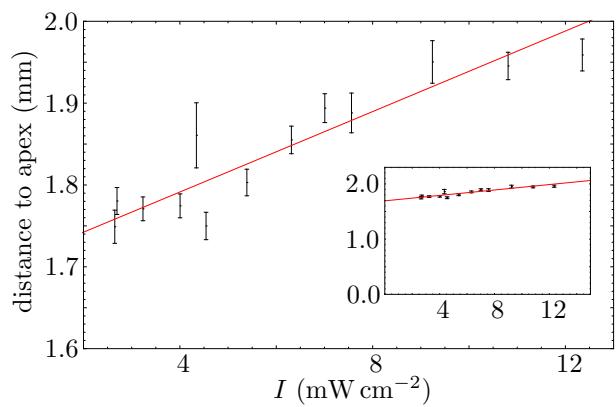}
\end{psfrags}
\caption{Distance of the MOT from the apex of the pyramid \textit{versus} intensity of the incident laser beam. As the intensity increases, the MOT moves away from the apex because the outward-going beam is more intense. Inset: The movement is small compared with the distance to the apex.
\label{fig:Imbalance}
}}
\end{figure}

In a standard 6-beam MOT, pairs of light beams propagate in opposite directions with equal intensity and opposite circular polarisation. This produces balanced forces that place the MOT at the centre of the magnetic quadrupole, where the magnetic field is zero, regardless of the intensity of the light. By contrast, the radiation pressure in our $70.5^\circ$ pyramid is unbalanced vertically at zero magnetic field, due to mismatches in both intensity and polarisation of the beams. We therefore expect the equilibrium position of the MOT to move up and down on the pyramid axis when the intensity of the light is varied. This behaviour is demonstrated by the data in \fig{fig:Imbalance}(a). As the intensity increases, the MOT moves further from the apex, indicating that the outward-going light is more intense than the incident beam, despite the fact that it has been reflected twice. This is due to a concentration of the reflected light into a smaller cross sectional area, as noted in \cite{trupke06}.  Over a wide range of intensities, above and below the intensity that optimises the MOT, we see a movement that is linear in intensity with slope $24.5\,\mu$m/(mW/cm$^{2}$). As shown by the inset in \fig{fig:Imbalance}, this movement is small compared with the total distance to the apex.

\subsection{\label{sec:understanding}Understanding the dependence on detuning and intensity}
\begin{figure}[b]
{\footnotesize
\centering
\subfigure{
\begin{psfrags}
\psfrag{S0}[tc][tc]{ $0.0$}%
\psfrag{Sm1252}[tc][tc]{ $~$}%
\psfrag{Sm12}[tc][tc]{ $-10$}%
\psfrag{Sm152}[tc][tc]{ $-15$}%
\psfrag{Sm1752}[tc][tc]{ $~$}%
\psfrag{Sm22}[tc][tc]{ $-20$}%
\psfrag{Sm251}[tc][tc]{ $~$}%
\psfrag{Sm51}[tc][tc]{ $-5$}%
\psfrag{Sm751}[tc][tc]{ $~$}%
\psfrag{W0}[cr][cr]{ $0$}%
\psfrag{W11}[cr][cr]{ $1.0$}%
\psfrag{W2}[cr][cr]{ $0.2$}%
\psfrag{W4}[cr][cr]{ $0.4$}%
\psfrag{W6}[cr][cr]{ $0.6$}%
\psfrag{W8}[cr][cr]{ $0.8$}%
\psfrag{X}[tc][tc]{ $\delta/2\pi$ (MHz)}%
\psfrag{Y}[bc][bc]{ Signal (Arb. Un.)}%
\psfrag{a}[br][br]{ \large $(a)$}%
\includegraphics[width=0.45\columnwidth]{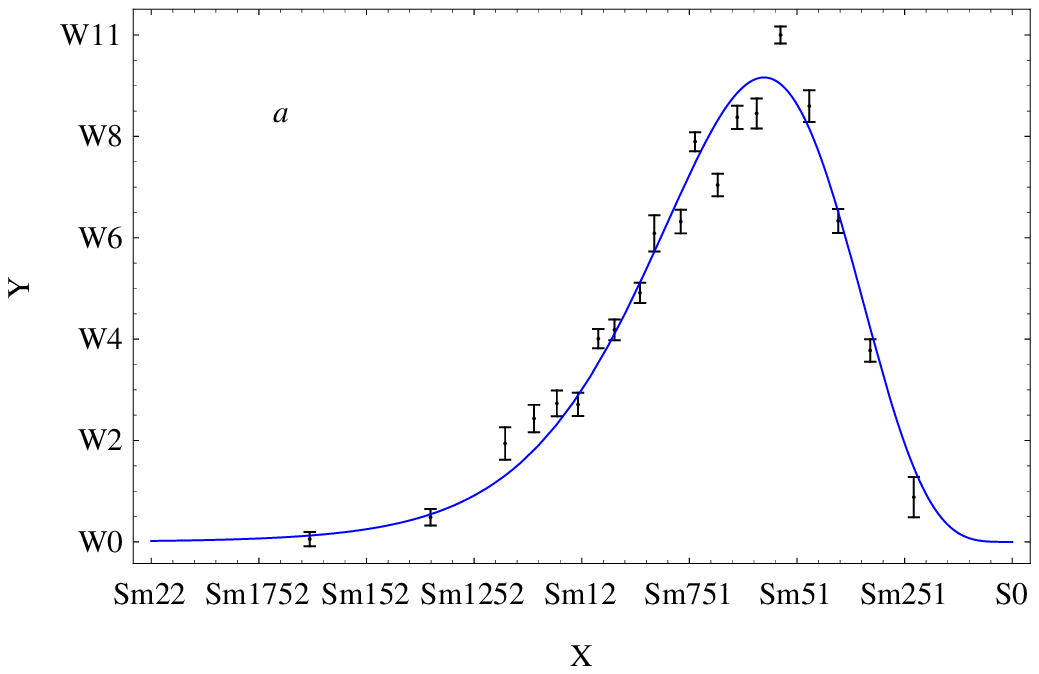}
\end{psfrags}
}
\hspace{0.5cm}
\subfigure{
\psfrag{S1252}[tc][tc]{ $~$}%
\psfrag{S12}[tc][tc]{ $10$}%
\psfrag{S152}[tc][tc]{ $15$}%
\psfrag{S1752}[tc][tc]{ $~$}%
\psfrag{S22}[tc][tc]{ $20$}%
\psfrag{S251}[tc][tc]{ $~$}%
\psfrag{S51}[tc][tc]{ $5$}%
\psfrag{S751}[tc][tc]{ $~$}%
\psfrag{W0}[cr][cr]{ $0.0$}%
\psfrag{W11}[cr][cr]{ $1.0$}%
\psfrag{W2}[cr][cr]{ $0.2$}%
\psfrag{W4}[cr][cr]{ $0.4$}%
\psfrag{W6}[cr][cr]{ $0.6$}%
\psfrag{W8}[cr][cr]{ $0.8$}%
\psfrag{b}[bl][bl]{ \large (b)}%
\psfrag{X}[tc][tc]{ $I$ (mW\,cm$^{-2}$)}%
\psfrag{Y}[bc][bc]{ Signal (Arb. Un.)}%
\includegraphics[width=0.45\columnwidth]{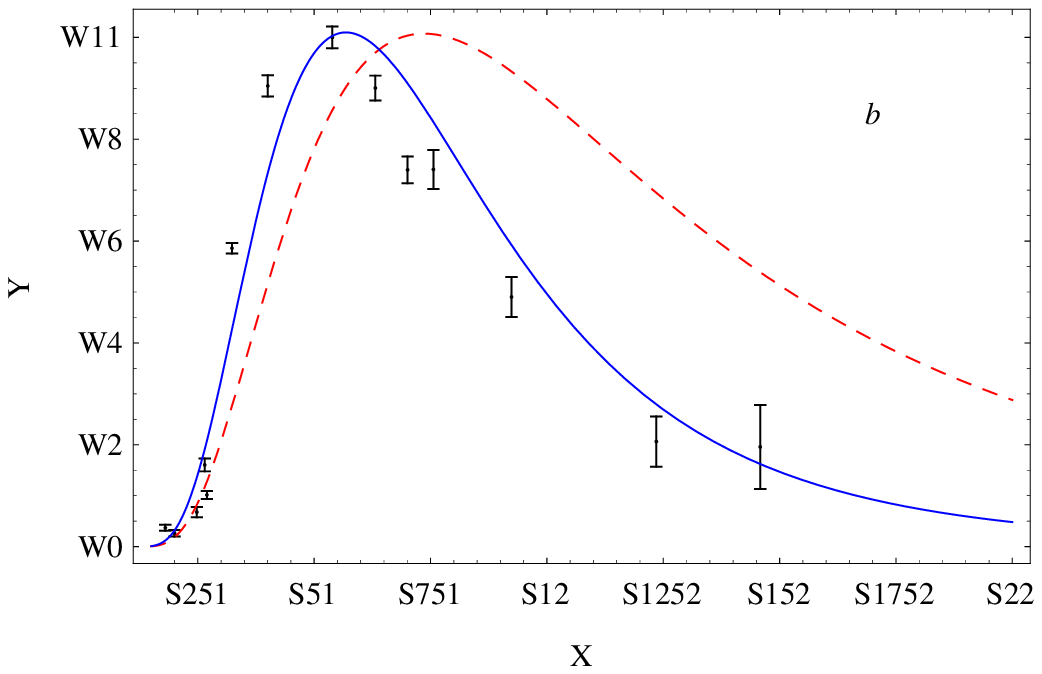}
}
\caption{Optimising the MOT fluorescence. Each data point represents the average of several experiments. (a) Variation with detuning for an incident intensity of $5\,\mbox{mW/cm}^2$. The line is the normalised function $f$ described in the text. (b) Variation with incident laser intensity at a detuning of $-10\,$MHz. The dashed (red) line is the function $f$. After correcting for the effect of cloud movement on the fluorescence intensity we obtain the solid (blue) curve.\label{fig:atomnumbers}}
}
\end{figure}

Figure~\ref{fig:atomnumbers}(a) shows the integrated count over the image of the MOT in an $L=4.2\,$mm pyramid \textit{versus} the detuning $\delta$ of the laser from resonance on the $F=3\rightarrow4$, D2 transition of $^{85}$Rb at a wavelength of $\lambda = 780\,$nm. The intensity $I$ of the incident laser beam is $5\,\mbox{mW/cm}^2$, while the repump intensity is $1\,\mbox{mW/cm}^2$ (the atom number is largely insensitive to repump intensity once it exceeds $0.5\,\mbox{mW/cm}^2$). We see a peak in the fluorescence, centred at $\delta/2\pi = -6\,$MHz and having a width (FWHM) of $\sim5.5\,$MHz. This peak represents the optimum conditions for the MOT and corresponds to having $7\times 10^3$ atoms in the cloud.

In order to model this behaviour, we consider that the signal should be proportional to the number of atoms in the MOT, $N_{\infty}$, and to the scattering rate per atom $S\propto I/(1+(\delta/\gamma)^2+\xi I)$. Here $\gamma=3\,$MHz is half the natural linewidth and $\xi I$ is the saturation parameter that accounts for power broadening. The number $N_{\infty}$ is proportional to $v_c^4$ (see Eq.~(\ref{eq:N})) and hence to $\alpha^4$, where $\alpha$ is the coefficient of friction. For this scattering rate $S$, the corresponding Doppler-cooling friction coefficient is $\alpha\propto I\delta /(1+(\delta/\gamma)^2+\xi I)^2$. Thus we compare our data with a function $f$ of the form $N_{\infty}S\propto f= I^5\delta^4 /(1+(\delta/\gamma)^2+\xi I)^9$. The data points in Fig.~\ref{fig:atomnumbers}(a) are well described by this function with the choice $\xi=2.4$, together with a suitable over-all vertical scaling.

The expected value of the coefficient $\xi$ can be estimated by considering the intensity of the light at the position of the MOT together with its polarisation and the corresponding saturation intensity. The total intensity is roughly $6I$ and the polarisation is somewhere between pure circular ($I_{sat}=1.7\,\mbox{mW/cm}^2$)  and isotropic ($I_{sat}=3.9\,\mbox{mW/cm}^2$). Thus, we can reasonably expect $\xi$ to be in the range $(1.5 - 3.5)\,\rm{mW}^{-1}\rm{cm}^2$, which is entirely consistent with the value that we derive from Fig.~\ref{fig:atomnumbers}(a). Since we have used the Doppler friction here, the agreement suggests that Doppler cooling is an adequate description for this MOT and that the Sisyphus mechanism is not playing an important role. One possible reason for the absence of appreciable sub-Doppler cooling is the displacement of the cloud from zero magnetic field, which is discussed in the previous section.

Figure~\ref{fig:atomnumbers}(b) shows the fluorescence of the same MOT, measured \textit{versus} incident intensity $I$, with a detuning of $\delta/2\pi=-10\,$MHz. The dashed (red) line shows the same function $f$ with the same $\xi=2.4$ and no free parameters, except for the over-all scaling. Although this curve captures the general trend, it overestimates the signal at high intensity. That is because the MOT fluorescence also changes with the position of the cloud (Sec.~\ref{sec:positiondependence}) and the position changes with the incident intensity (Sec.~\ref{sec:intensitydependence}). On multiplying $f$ by the Lorentzian representing the intensity dependence due to movement of the cloud, we obtain the solid (blue) line in Fig.~\ref{fig:atomnumbers}(b), which is in good agreement with the data.

\subsection{\label{subsec:ScalingLaws}Scaling of atom number with pyramid size}

\begin{figure}[h!]
\centering
{\footnotesize
\begin{psfrags}
\psfrag{N}[cc][cc]{Atom number}
\psfrag{X}[cc][cc]{$\sqrt{\mathrm{Area}}$\,(mm)}
\psfrag{a}[cc][cc]{$10^{6}$}
\psfrag{b}[cc][cc]{$10^{5}$}
\psfrag{c}[cc][cc]{$10^{4}$}
\psfrag{d}[cc][cc]{$10^{3}$}
\psfrag{e}[cc][cc]{$10^{2}$}
\psfrag{f}[cc][cc]{$10^{1}$}
\psfrag{g}[cc][cc]{$3$}
\psfrag{h}[cc][cc]{$4$}
\psfrag{i}[cc][cc]{$5$}
\psfrag{j}[cc][cc]{$6$}
\psfrag{k}[cc][cc]{$7$}
\psfrag{l}[cc][cc]{$8$}
\psfrag{m}[cc][cc]{$9$}
\psfrag{n}[cc][cc]{$10$}
\psfrag{o}[cc][cc]{\scriptsize $10$}
\psfrag{p}[cc][cc]{\scriptsize $20$}
\psfrag{q}[cc][cc]{\scriptsize $-20$}
\psfrag{r}[cc][cc]{\scriptsize $-10$}
\psfrag{s}[cc][cc]{\scriptsize $0.1$}
\psfrag{t}[cc][cc]{\scriptsize $-0.1$}
\psfrag{v}[cc][cc]{$v$\,(ms$^{-1}$)}
\psfrag{F}[cc][cc]{$F\,(\hbar k \gamma)$}
\includegraphics[width = 0.75\columnwidth]{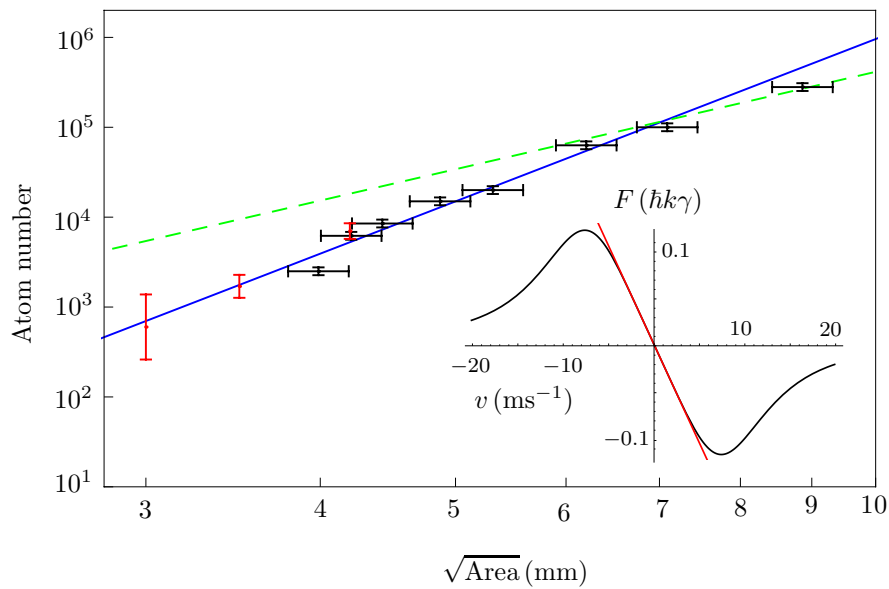}
\end{psfrags}
\caption{Scaling law. Red and black data points correspond to MOTs formed in silicon pyramids and macroscopic glass pyramids respectively. For pyramids below $\sim6\,$mm we find a scaling well described by the relation $N_{\infty} \propto L^{6}$ illustrated by the solid (blue) line. The dashed (green) line is the empirical scaling law for large MOTs, $N_{\infty} \propto L^{3.6}$. Inset: Slowing force experiences by atoms {\em versus} their velocity. The red line shows the approximation $F = -\alpha v$. \label{fig:scaling}
}}
\end{figure}

Throughout this paper we have made use of the scaling law in Eq.~\ref{eq:N} that $N_{\infty}\propto L^2 v_c^4$. We have further assumed that the MOT is operating within the linear part of the friction curve, shown inset in \fig{fig:scaling}, where the frictional force on an atom is proportional to its velocity with coefficient $\alpha$. This gives a capture velocity proportonal to the pyramid size $L$ and hence $N_{\infty}\propto L^6$. In order to check this, we measured the  MOT fluorescence using a range of pyramids having different opening apertures. The smallest pyramids were etched in silicon wafers and had openings of $L = 4.2, 3.5$ and $3\,$mm. We did not make the measurement in pyramids smaller than this as the MOT was hard to see and difficult to optimise. The etching method was not suitable for making larger pyramids because of the very thick wafers and unreasonably long etching times required. In order to extend the range to larger apertures, we used an $L = 16\,$mm glass pyramid with the same $70.5^{\circ}$ geometry and Al coating. An adjustable circular aperture set the diameter of the beam entering this larger pyramid to achieve a range of effective sizes, where the effective value of $L$ is taken to be the square root of the aperture area.

The results are shown in Fig.~\ref{fig:scaling} over the range $L = 3-9\,$mm, for an incident laser intensity of $5\,\mbox{mW/cm}^2$ and a detuning of $-8\,$MHz. We see that up to an aperture size of $\sim6\,$mm, the atom number does indeed follow the $L^6$ scaling law indicated by the solid (red) line. For larger pyramids, the number increases more slowly, and by the time the aperture is as large as $1\,$cm, the scaling is closer to the empirical $L^{3.6}$ scaling recommended in \cite{lindquist92}. This can be understood by considering the capture velocity. Scaling linearly from our measurement in Sec.~\ref{sec:loadloss}, a 6\,mm pyramid has a capture velocity of $\sim 7\,$m/s. But, as shown in the inset in Fig.~\ref{fig:scaling} this corresponds to the velocity at which the frictional force reaches its maximum. Roughly speaking the damping force is linear for velocities up to $\delta \lambda$, where the Doppler shift is equal to the detuning, beyond which the force decreases. Hence, the capture velocity in our experiments grows linearly up to a pyramid size of $\sim 6\,$mm, above which it grows ever more slowly.

\section{\label{sec:summary}Summary and outlook}

We have investigated the performance of small pyramid MOTs that are etched into a silicon wafer. We have shown how the atom cloud can be imaged against the background of scattered light and have used these images to measure the loading and loss rates of the trap. By moving the atom cloud within the pyramid, we have measured how the number of trapped atoms is affected by proximity to the walls and have shown that the main effect is the loss of cold atoms from the tail of the Gaussian density distribution. We have studied how the MOT fluorescence varies with position of the cloud, intensity of the light and detuning, and we find that the behaviour can be understood in detail with the help of a simple Doppler cooling model. The scaling of atom number with MOT size has also been understood. This constitutes the first quantitative study of MOT behaviour in a small confined space.

Miniature pyramids show promise as a chip-scale source of cold atoms for integration into other experiments because they are simple and robust, requiring only one circularly polarised laser beam and essentially no alignment. For future applications, it would be helpful to eliminate the imbalance of beams so that the MOT position remains fixed at the zero of magnetic field as the laser intensity varies. This should also lead to lower MOT temperatures through the operation of the sisyphus mechanism. Such a balance might be achieved by judicious choice of wall coating to lower the intensity of the reflected beams. Alternatively, it would be promising if $90^{\circ}$ pyramids could be fabricated, since these are automatically balanced when the walls have high reflectivity~\cite{resnik05,footnote2}.

Many experiments require low pressures ($\sim10^{-11}$\,mbar) to ensure that the ultracold atoms are undisturbed by background gas collisions for many seconds. In such cases, the pyramid MOT can be incorporated into a double-sided chip, where both the front and back are used. A small vacuum chamber partitioned by the chip would have Rb pressure on the pyramid side. Atoms would pass from the MOT to the low pressure side through a small aperture etched into the apex of the pyramid, much as in an LVIS source \cite{lu96,kohel03}. On arrival, the atoms are trapped and controlled by the usual microfabricated wires and optical elements \cite{reichelbook}.

The integrated pyramid MOT is also a promising source of small atom clouds, or even single atoms, without the need for extremely high magnetic field gradients \cite{frese00}. From our scaling law studies we conclude that a pyramid with a $\sim1$\,mm aperture would trap a single atom.

\section{Appendix}
In this appendix we derive the expression for the capture rate $R$ used in Sec.~\ref{sec:loadloss}.
In the Rb vapour, the number density of atoms whose velocity lies within solid angle $d\Omega$ is $n_{vap} \frac{d\Omega}{4 \pi}$. The Maxwell-Boltzmann probability of a speed between $v$ and $v + dv$ is $f(v) dv = \frac{4}{\sqrt{\pi}} \frac{v^2}{\beta^{3}} e^{-v^2/\beta^{2}}dv$, where $\beta = \sqrt{2k_{B}T/m}$ is the  most probable speed. Hence the flux of atoms moving in this speed range within this solid angle is
\begin{equation}
d\Phi = n_{vap} \frac{d\Omega}{4 \pi} v f(v) dv.
\end{equation}
The number of these per second striking an annulus between $b$ and $b+db$ is $2\pi b\,db\,d\Phi$. Let us assume a spherical capture volume with cross sectional area $A$ and radius $a = \sqrt{A/\pi}$. Integrating over all solid angles and over all speeds up to the capture velocity $v_c(b)$ and over all impact parameters up to $a$, the capture rate is
 \begin{equation}
R = \int_{0}^{a} 2 \pi b db \int_{0}^{v_c(b)} n_{vap} v f(v) dv\,.
\label{Eq:R1}
\end{equation}
Ignoring any deflection of the atoms, the distance available for stopping is $2\sqrt{a^2-b^2}$. Since the capture velocity is proportional to this distance, $v_{c}(b) = v_c\sqrt{1-(b/a)^2}$, where $v_c$ is the maximum capture velocity. After performing the integral in Eq.~(\ref{Eq:R1}) and making the approximation  $(v_{c}/\beta)^2\ll 1$, we obtain the capture rate
\begin{equation}
 R =  \frac{8 A v_{c}^{4}}{3 \pi^{2} \bar{v}_{vap}^{3}} n_{vap}\,,
\end{equation}
where the mean speed $\bar{v}_{vap}$ is equal to $\frac{2}{\sqrt{\pi}}\beta$.

\section{Acknowledgments}
This work is supported by the CEC Seventh Framework projects 216744 (CHIMONO) and 247687 (AQUTE), the UK EPSRC and the Royal Society. FRM acknowledges support from the Mexican CONACYT and SEP programmes. We are grateful to J. Dyne and S. Maine for technical assistance.

\section*{References}
\bibliographystyle{unsrt}

\begin{thebibliography}{15}
\bibitem{footnote2}
We have recently been made aware that the group of J. Kitching at NIST in
  Boulder are attempting to microfabricate 1\,mm Si pyramids with an apex angle
  of $90^{\circ}$.
\end{thebibliography}

\end{document}